\documentclass[english]{lni}
\usepackage{tabularx}
\usepackage{graphicx} 
\usepackage{wrapfig} 
\begin{document}
%%% Mehrere Autoren werden durch \and voneinander getrennt.
%%% Die Fußnote enthält die Adresse sowie eine E-Mail-Adresse.
%%% Das optionale Argument (sofern angegeben) wird für die Kopfzeile verwendet.
\title[RKG]{Research Knowledge Graphs in NFDI4DataScience: Key Activities, Achievements, and Future Directions}
%% \subtitle{Untertitel / Subtitle} % if needed
\author[1]{Kanishka Silva}{kanishka.silva@gesis.org}{0000-0003-2958-9552}
\author[2]{Marcel R. Ackermann}{marcel.r.ackermann@dagstuhl.de}{0000-0001-7644-2495}%
\author[3]{Heike Fliegl}{Heike.Fliegl@fiz-Karlsruhe.de}{0000-0002-7541-115X}
\author[3]{Genet-Asefa Gesese}{genet-asefa.gesese@fiz-karlsruhe.de}{0000-0003-3807-7145}
\author[4]{Fidan Limani}{f.limani@zbw.eu}{0000-0002-5835-2784}
\author[1]{Philipp Mayr}{philipp.mayr@gesis.org}{0000-0002-6656-1658}
\author[1]{Peter Mutschke}{peter.mutschke@gesis.org}{0000-0003-3517-8071}
\author[5]{Allard Oelen}{allard.oelen@tib.eu}{0000-0001-9924-9153}
\author[1]{Muhammad Asif Suryani}{asif.suryani@gesis.org}{0000-0003-1669-5524}
\author[1]{Sharmila Upadhyaya}{sharmila.upadhyaya@gesis.org}{0009-0003-7142-3887}
\author[1]{Benjamin Zapilko}{benjamin.zapilko@gesis.org}{0000-0001-9495-040X}
\author[3]{Harald Sack}{harald.sack@fiz-karlsruhe.de}{0000-0001-7069-9804
}%{}
\author[1,6]{Stefan Dietze}{stefan.dietze@hhu.de}{0009-0001-4364-9243}
 
\affil[1]{GESIS -- Leibniz Institute for the Social Sciences\\Cologne\\Germany}
\affil[2]{DBLP computer science bibliography\\Schloss Dagstuhl - LZI\\Trier\\Germany}
\affil[3]{FIZ Karlsruhe -- Leibniz Institute for Information Infrastructure GmbH\\Eggenstein-Leopoldshafen\\Germany}
\affil[4]{ZBW -- Leibniz Information Centre for Economics\\Kiel\\Germany}
\affil[5]{TIB -- Leibniz Information Centre for Science and Technology\\Hannover\\Germany}
\affil[6]{Heinrich Heine University Düsseldorf\\Düsseldorf\\Germany}

\maketitle

\begin{abstract}
As research in Artificial Intelligence and Data Science continues to grow in volume and complexity, it becomes increasingly difficult to ensure transparency, reproducibility, and discoverability. 
To address these challenges, as research artifacts should be understandable and usable by machines, the NFDI4DataScience consortium is developing and providing Research Knowledge Graphs (RKGs).
Building upon earlier works, this paper presents recent progress in creating semantically rich RKGs using standardized ontologies, shared vocabularies, and automated Information Extraction techniques.
Key achievements include the development of the NFDI4DS ontology, metadata standards, tools, and services designed to support the FAIR principles, as well as community-led projects and various implementations of RKGs.
Together, these efforts aim to capture and connect the complex relationships between datasets, models, software, and scientific publications.
\end{abstract}
\begin{keywords}
NFDI4DS \and Research Knowledge Graphs \and FAIR Principles \and Information Extraction \and Research Knowledge Graph Integration
\end{keywords}
%%% Beginn des Artikeltexts
\section{Introduction}
Research outputs in Artificial Intelligence (AI) and Data Science (DS) have significantly increased in volume and complexity.
Related datasets and models are updated, extended, derived from, or combined with other artifacts in this context.
Without adequate representations of these varieties and intricate dependencies, it is difficult to ensure transparency, reproducibility, reuse, and discoverability~\cite{matthaus_zloch_research_2025}.
A state-of-the-art method for representing scholarly entities and their relationships in a standardized, machine-actionable way is Research Knowledge Graphs (RKGs).
The German National Research Data Infrastructure for Data Science and Artificial Intelligence (NFDI4DS\footnote{\url{https://www.nfdi4datascience.de/}}) aims to address the aforementioned challenges through a joint RKG ecosystem that provides efficient access to AI and DS resources, most notably, data, code, publications, and Machine Learning~(ML) models, as well as transparent information on their interdependencies~\cite{karmakar2023}.
This paper\footnote{An updated overview building on the frameworks of \Citet{karmakar2023} and \Citet{matthaus_zloch_research_2025}, shifting from initial RKG vision within NFDI4DS to recent developments and future directions.} focuses on presenting a more recent account of key components, achievements, and future directions for extracting and linking AI and DS research artifacts into coherent RKGs.

\section{Background and Related Works}
RKGs such as those described in \Citet{Jaradeh2019,Ackermann2024,Sinha2015,Danilo2020} capture detailed, machine-actionable, semantically rich, interconnected descriptions of scholarly entities.
Using a formal and structured data model, RKGs represent various research-related entities such as publications, datasets, software tools, authors, ML models, scientific tasks, and evaluation metrics.
These entities are encoded in a graph-based structure, facilitating complex queries across (linked) research data.
This structure supports retrieval of contextual and provenance information -- key elements for ensuring transparency, reproducibility, and accountability in research processes~\cite{matthaus_zloch_research_2025}.
Moreover, RKGs help implement FAIR (Findable, Accessible, Interoperable, and Reusable) principles in research infrastructures~\cite{Wilkinson2016}.
These are achieved by (1) integrating persistent identifiers (PIDs), such as DOIs and ORCIDs; (2) including semantic terminologies to describe their resources; and (3) providing standards-based access capabilities like SPARQL.
Further, RKGs provide essential functionalities to structure and align interoperable metadata to improve the discoverability and re-usability.

\section{Scholarly Information Extraction and RKG Construction}
\begin{figure}
    \centering
    \includegraphics[width=0.6\linewidth]{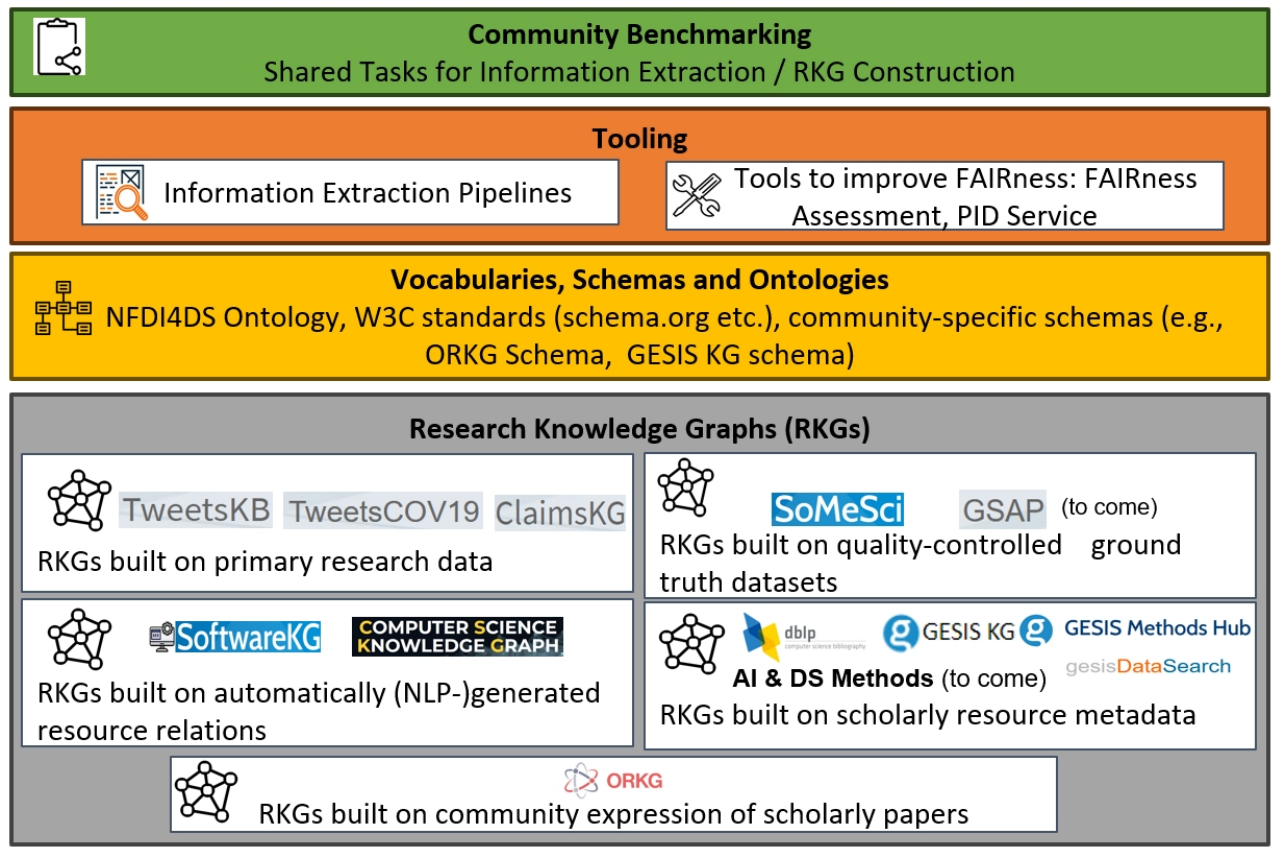}
    \caption{Service diagram layer structure}
    \label{fig:layer_struct}
\end{figure}

Within NFDI4DS, a set of components has evolved to support different life cycle phases of developing and reusing RKGs.
Decoupling the implementation for the different components enables one to address the corresponding requirements almost independently.
Thus, we can handle the inclusion of a new artifact in one of the RKGs by reusing or designing a best-performing Information Extraction~(IE) pipeline.
We have organized these components as follows: (1) community benchmarking; (2) tools and pipelines for IE and FAIR assessment; (3) vocabularies, schemas, and ontologies needed for RKG construction; (4) the RKGs themselves (See Fig. \ref{fig:layer_struct}).

Community benchmarking focuses around modeling the research artifacts of interest; a terminology component -- vocabularies, schema, ontologies, etc. -- forms the basis for describing these artifacts and is supported by dedicated services for ontology development, schema alignment, and metadata standardization; tooling provides methods and pipelines to extract the entities from collections (structured, or unstructured, like bibliographic and web resources), and assess their FAIRness; finally the RKGs are populated with the previous outcomes.

\subsection{Community Benchmarking}
The primary mechanism for actively engaging the community is through the organization of shared tasks\footnote{For instance, organized at venues such as Scholarly Document Processing (SDP) workshop \url{https://sdproc.org/2025/}, International Workshop on Semantic Evaluation (SemEval) \url{https://semeval.github.io/}, and ISWC Challenge Tracks \url{https://iswc2025.semanticweb.org}}, which have proven to be catalysts for advancing state-of-the-art methods~\cite{semeval, conll} in specific areas and bringing together disparate sub-communities~\cite{Filannino2018}.
Shared tasks related activities serve as vital benchmarks for the IE methods that we develop, hence they are fundamental for automatically constructing RKGs.
NFDI4DS established a consortium-wide working group on \emph{Shared Tasks}\footnote{\url{https://www.nfdi4datascience.de/community/shared-tasks/}}, which specifically coordinates these activities, overseeing tasks that generate the structured information and metadata necessary for populating RKGs.
As a key activity which coordinates community-driven evaluation tasks to benchmark IE methods, two editions of the Software Mention Detection (SOMD) shared task has been organized: SOMD-2024~\cite{Frank2024}, held at the Natural Scientific Language Processing Workshop at ESWC 2024\footnote{\url{https://nfdi4ds.github.io/nslp2024/}} and SOMD-2025\footnote{\url{https://sdproc.org/2025/somd25.html}} hosted by the Scholarly Document Processing Workshop at ACL 2025. These efforts demonstrate how structured community benchmarks contribute to the development of high-quality RKGs within NFDI4DS.

\subsection{Tooling}
The \emph{Tooling} component is dedicated to supporting various pipelines and services essential for creating, enriching, and managing RKGs. 
It encompasses functionalities such as IE pipelines for populating RKGs and tools designed to enhance the FAIR principles of research artifacts.
The development and evaluation of these tools, including research on scholarly IE and RKG linking, contribute to a more interconnected and machine-actionable landscape for AI and DS related digital artifacts.

\textbf{IE Pipelines.}
Research artifacts are distributed across various open repositories like Hugging Face, arXiv, GitHub, and Zenodo. 
Furthermore, Open Science practices have empowered research communities by democratizing access to these artifacts, resulting in a wide range of heterogeneous applications, domains, and diverse metadata features, which adds to the requirements the RKGs have to address.
Targeting a growing landscape of research artifacts across open repositories, multiple pipelines have been developed to extract relevant metadata features.
For instance, metadata features aim to cover provenance, which is crucial to capture as these metadata features tend to evolve over time~\cite{suryani2025modelcard, SuryaniKM24}, licensing, which is important during the reuse of an artifact (and is also reflected in its FAIR assessment), or other descriptive elements, such as number of downloads, stars, and fork (GitHub), or the ML model card (Hugging Face).
In terms of execution time, these pipelines have to be able to perform at scale, considering that repositories such as the ones just mentioned contain large collections.

\textbf{FAIR Assessment.}
The adoption of FAIR principles provides greater reuse of digital resources, regardless of the domain.
Thus, exploring their adoption, including corresponding assessments, is another set of services of the \emph{Tooling} component.
RKGs, by design, support some of the FAIR principles and quality assurance standards for research artifacts.
Consider persistent identifiers, typified entities, or provision of rich metadata description (for Findability), available based on standard protocols (for Accessibility), described via community standards (ontologies) -- for Interoperability, to name but a few, attest to this.
However, to do so, we have to specify the FAIR adoption and assessment practices for the artifacts of interest.
For some, such as datasets and research software, we explored existing FAIR assessment initiatives~\cite{RDA2020FAIR, Devaraju2022FAIRsFAIR, ChueHong2022FAIR4RS}, while for others, such as ML models, we extended relevant existing ones and proposed a new framework~\cite{Limani2024FAIRML}.
We explored the applicability of Large Language Models (LLMs) as implementation means for FAIR evaluation by applying LLM models for automatically assessing FAIR metric tests\footnote{See the tool for dataset FAIR assessment employing LLMs at \url{https://data-science.hsnr.de/fairway/assess}}.
Finally, FAIR assessment provides a quality assurance dimension of the artifacts with which we populate the RKGs.

\subsection{Vocabularies, Schemas and Ontologies}
Structured vocabularies, schemas, and ontologies are essential to improve reproducibility and transparency in AI and DS. The NFDI4DS metadata working group\footnote{\url{https://www.nfdi.de/section-meta/task-force-metadata}} was established to explore existing metadata standards and schemas, and identify and address any schema-related gaps for the community.
Moreover, the activities within this group involve working with foundational ontologies to ensure interoperability within NFDI.
The NFDICore Ontology~\cite{bruns2024nfdicore20bfocompliantontology}, for example, is a mid-level ontology that models metadata for NFDI resources and connects them to foundational ontologies like BFO and standards such as schema.org\footnote{\url{https://schema.org/}} to support interoperability across consortia.
Furthermore, the NFDI4DS Ontology~\cite{gesese2024nfdi4dsobfocompliantontology} (in preparation) is a domain-specific, modular extension of NFDICore Ontology, providing the conceptual basis for the NFDI4DS Knowledge Graph.
Finally, recognizing the importance of standards for describing ML models and software, NFDI4DS is currently analyzing existing metadata schemas for ML models to develop or extend a suitable schema in the future.

\subsection{RKGs in NFDI4DS and Their Categorization}

\begin{table}[ht!]
    \centering
    \begin{tabularx}{\textwidth}{|X|X|}
         \hline
         \textbf{Category} & \textbf{RKGs}\\
         \hline

         Scholarly resource metadata & GESIS KG\footnotemark[16], dblp KG~\cite{Ackermann2024}, gesisDataSearchKG\footnotemark[17], MethodsHub KG (in preparation), KG on AI and DS Methods (in planning), Hugging Face KG (in preparation)\\
         \hline
         Quality-controlled ground truth dataset & SoMeSci~\cite{Schindler2021}, GSAP KG (in planning) \\
         \hline
         Primary research data  & TweetsKB~\cite{Fafalios2018}, TweetsCOV19~\cite{Dimitrov2020}, ClaimsKG~\cite{Tchechmedjiev2019} \\
         \hline
         Automatically-generated resource relations & SoftwareKG~\cite{Schindler2022}\\
         \hline
         Community expression of scholarly papers & ORKG~\cite{Jaradeh2019} \\
         \hline
         
    \end{tabularx}
    \caption{Categorization of RKGs in NFDI4DS}
    \label{tab:overview}
\end{table}

\footnotetext[16]{\url{https://search.gesis.org/research_data/SDN-10.7802-2878}}
\footnotetext[17]{\url{https://data.gesis.org/gesisdatasearchkg}}

A major focus of NFDI4DS is creating and populating RKGs relevant to the AI and DS community.
Considering the variety of research artifacts in this domain, we create and populate RKGs dedicated to the entities of a certain artifact we target, including the use cases.
Thus, the RKGs we provide fall into five distinct categories~\cite{matthaus_zloch_research_2025} as in Table \ref{tab:overview}.

A key category of RKGs \textit{built on scholarly resource metadata} from bibliographic and research data repositories. 
These serve as a primary access point for researchers to search for publications and related artifacts.
They are built on stable schemas with a focused subset of vocabularies.
Below, we briefly describe these RKGs.

\begin{itemize}
    \item \textbf{dblp KG}~\cite{Ackermann2024} offers a semantic view of all entities and relationships within the DBLP Computer Science Bibliography\footnote[18]{\url{https://dblp.org}}.
    It contains over 510 million RDF statements about more than 3.7 million authors, 7.8 million publications, and 9,000 publication venues. What sets dblp KG apart is its emphasis on manual curation and quality-controlled entity disambiguation and linkage.
    A public SPARQL query service\footnote[19]{\url{https://sparql.dblp.org}} is available to the community and provides data synchronized with the current DBLP database. Monthly dump snapshots of the KG~\cite{dblp_rdf_ntriples} are published under the CCO open data license.

    \item \textbf{GESIS KG} represents metadata on social science research resources available through GESIS Search\footnote[20]{\url{https://search.gesis.org/}}.
    It emphasizes the semantic relationships between different resources such as datasets, publications, and survey variables.
    By offering flexible access methods (SPARQL endpoints, OAI-PMH API), it allows it to be incorporated in other systems, e.g., into the NFDI4DS Gateway and the KGI4NFDI basic service registry, which makes it accessible for discovery and federation across RKGs in the NFDI.
    The OAI-PMH API is being extended to connect GESIS KG to OpenAIRE, and through it, to the European Open Science Cloud (ESOC), which could also serve as a blueprint for connecting other NFDI data points with ESOC.

    \item \textbf{gesisDataSearchKG} provides a comprehensive data model designed to represent the metadata of social science datasets harvested from various repositories.
    This resource represents various artifacts of social science datasets using available vocabularies, such as DDI Alliance\footnote[21]{\url{https://ddialliance.org/}}, DataCite\footnote[22]{\url{https://datacite.org/}}, and schema.org, supporting interoperability and semantic representation of dataset metadata.

    \item \textbf{KG of AI and DS methods} is being built to assist researchers in finding and applying AI and DS methodologies for their data.
    As an initial step, the input data for this KG will be sourced from the GESIS MethodsHub KG, with a strategy for expansion through partner contributions across the NFDI4DS consortium and other collaborations.
    This RKG is planned to connect with the GESIS KG.

    \item \textbf{NFDI4DS KG} consists of two main components: (1) Research Information Graph (RIG); and (2) Research Data Graph (RDG).
    The former included metadata about the NFDI4DS consortium's resources, persons, and organizations as described above, while the latter covers content-related data from the consortium's heterogeneous data sources.
    RIG will serve as the backend for the NFDI4DS web portal, facilitating interactive data access and management.
    Both RIG and EDG will be accessible and searchable via the NFDI4DS portal.
    The first version of the NFDI4DS KG\footnote[23]{\url{https://nfdi.fiz-karlsruhe.de/4ds/sparql}}\footnote[24]{\url{https://nfdi.fiz-karlsruhe.de/4ds/shmarql}} in the form of the RIG is publicly available, and data browsing using SPARQL queries as well as SHMARQL\footnote[25]{\url{https://shmarql.com/}} is available.

    \item  \textbf{Hugging Face KG (HFKG)} is currently under development, addressing the rapid growth and advancements in ML models at Hugging Face\footnote[26]{\url{https://huggingface.co/models}}. 
    HFKG targets heterogeneous metadata features such as models, datasets, libraries, licenses, and tasks, capturing their relationships and enabling linkages to external resources such as authors with models~\cite{suryani_2025_15125488, suryani2025modelcard}.
    HFKG also aims to address reasoning, searching, and exploration across the ML ecosystem by incorporating temporal and provenance information to trace the evolution of models and related artifacts.
    These features position HFKG as a tool that not only supports metadata-driven exploration, such as identifying popular architectures and license distributions, but also opens opportunities for assessing reproducibility practices and scholarly linkages, ultimately contributing to the digitalization of scientific infrastructures and improved interoperability across research communities.
\end{itemize}

RKGs \textit{build upon a quality-controlled ground truth dataset} that provides curated, annotated data for benchmarking.
Typically, smaller in scale, these RKGs employ a fixed, stable schema and a well-defined and focused subset of vocabulary.
A few examples follow.

\begin{itemize}
    \item \textbf{SoMeSci KG} is a gold standard KG of software mentions in scientific articles. 
    It contains high-quality annotations of 3,756 software mentions in 1,367 PubMed Central articles.

    \item \textbf{GSAP KG (GESIS Scholarly Annotation Project)\footnote[27]{\url{https://data.gesis.org/gsap/gsap-ner/}}} focuses on representing ML model and dataset entities, as they are explicitly mentioned in a manually annotated \emph{gold set} of 100 scholarly articles from the Computer Science domain~\cite{otto_gsap-ner_2023}.
    A key objective of this RKG is to capture the relationships that exist between identified ML models and dataset entities within the scholarly text.
    An ongoing task is to uncover GSAP entities and their relationships from large document corpora.
\end{itemize}

A third category of RKGs consists of those \textit{built on primary research data} collected by researchers seeking research data relevant to their discipline. 
These RKGs can vary in scale, potentially up to a larger scale, but typically involve a fixed and stable schema with a focused subset of vocabulary.

\begin{itemize}
    \item \textbf{TweetsKB} is a KG that includes metadata about 3.1 billion tweets (Feb. 2013 - Jun. 2023) and serves as a resource for social science research.
    Using IE methods, sentiments, entities, hashtags, and user mentions were extracted and published as Linked Data through a structured RDF schema.

    \item \textbf{TweetsCOV19} is a semantically annotated corpus of Tweets about the COVID-19 pandemic.
    It is a subset of TweetsKB consisting of 41,307,082 tweets posted between October 2019 and August 2022.

    \item \textbf{ClaimsKG} contains claims and their evaluation from fact-checking websites and links relevant entities with concepts in DBpedia.
    The latest release of ClaimsKG covers 74,066 claims and 72,128 claim reviews published between 1996 and 2023.
\end{itemize}

A fourth category of RKGs is \textit{built upon automatically generated resource relations}, constructed through NLP-based extraction from unstructured scholarly artifacts.
These RKGs consist of a fixed, stable schema and a focused subset of vocabulary, but often with potentially evolving datasets that could be scaled significantly.

\begin{itemize}
    \item \textbf{SoftwareKG} contains information about software mentions from more than 51,000 scientific articles from the social sciences, enabling analysis on the provenance of the research results, software citation analysis, as well as an assessment of the state and the role of open source software in science.
\end{itemize}

The final category of RKGs is \textit{built upon community expressions of scholarly papers}, collaboratively developed to enhance data discoverability.
These RKGs typically have an openly evolving schema with templates; vocabulary is recommended but not strictly enforced.
A prominent example within NFDI4DS is the ORKG

\begin{itemize}
    \item \textbf{ORKG}\footnote[28]{\url{https://orkg.org/data}}~\cite{Jaradeh2019} represents structured scholarly content extracted from scientific articles, specifically on research contributions, in addition to describing the paper's metadata.
    ORKG supports knowledge discovery and downstream features: (1) leaderboards; (2) tabular overviews of scholarly literature (ORKG Comparisons); (3) community-maintained live literature reviews (ORKG Reviews), and (4) AI-assisted literature search (ORKG Ask)\footnote[29]{\url{https://ask.orkg.org/}}~\cite{oelen2024orkg}.
\end{itemize}

\section{Conclusion and Future Directions}
In recent years, NFDI4DS has brought together diverse community efforts to construct and enrich RKGs by developing shared ontologies, building robust IE pipelines, and coordinating shared tasks to model research artifacts.
These activities have resulted in notable achievements, including creating interoperable tools, curated RKGs, and services that enhance the FAIRness of research outputs.
One promising direction is harmonizing research artifacts across repositories by leveraging modular foundations of RKGs.
This effort aims to unify metadata from platforms like Hugging Face, GitHub, Zenodo, and arXiv, reflecting artifacts throughout the research life cycle.
Additionally, focusing on provenance enables the capture of relationships such as \texttt{isDerivedFrom}, \texttt{isAuthoredBy}, or \texttt{usesDataset} that support metadata enrichment, reproducibility, and impact tracking through the integration of external identifiers, like ORCID and ROR.
The key vision is to create a globally interconnected and machine-actionable scholarly knowledge infrastructure, capable of facilitating seamless access and understanding of research, both within the AI and DS domains, and beyond the NFDI4DS community. 
To further support cross-disciplinary research, part of this vision also includes supporting federated queries between RKGs in NFDI4DS and those of other NFDI consortia.

\section*{Acknowledgements}
This work has been funded by the Deutsche Forschungsgemeinschaft (DFG, German Research Foundation), NFDI4DS (Grant number 460234259).
The authors also acknowledge the data sources cited and the individuals involved in this work.

\printbibliography
\end{document}